\PassOptionsToPackage{table,xcdraw}{xcolor}
\documentclass[sigconf,nonacm]{acmart}

\AtBeginDocument{%
  }

\usepackage{graphicx}
\usepackage{tcolorbox}
\usepackage{longtable}
\usepackage{tikz}
\usepackage{array}
\usepackage{multirow}
\usepackage[table]{xcolor} 
\usepackage{supertabular}

\definecolor{post}{HTML}{BAC8D3}
\definecolor{training}{HTML}{FFF2CC}
\definecolor{pre}{HTML}{F9F6D1}

\definecolor{trainingdata}{HTML}{FAD7AC}
\definecolor{prompts}{HTML}{CCE5FF}
\definecolor{output}{HTML}{E1D5E7}
\definecolor{agents}{HTML}{B0E3E6}

\begin{document}

\title{SoK: The Privacy Paradox of Large Language Models: 
Advancements, Privacy Risks, and Mitigation} 

\author{Yashothara Shanmugarasa}
\email{yashothara.shanmugarasa@data61.csiro.au}
\orcid{0000-0002-6414-9416}
\affiliation{%
  \institution{CSIRO's Data61}
  \city{Sydney}
  \state{NSW}
  \country{Australia}
}

\author{Ming Ding}
\email{ming.ding@data61.csiro.au}
\orcid{0000-0002-3690-0321}
\affiliation{%
  \institution{CSIRO's Data61}
  \city{Sydney}
  \state{NSW}
  \country{Australia}
}

\author{M.A.P. Chamikara}
\email{chamikara.arachchige@data61.csiro.au}
\orcid{0000-0002-4286-3774}
\affiliation{%
  \institution{CSIRO's Data61}
  \city{Melbourne}
  \state{Victoria}
  \country{Australia}
}

\author{Thierry Rakotoarivelo}
\email{thierry.rakotoarivelo@data61.csiro.au}
\orcid{0000-0001-7698-6214}
\affiliation{%
  \institution{CSIRO's Data61}
  \city{Sydney}
  \state{NSW}
  \country{Australia}
}

\begin{abstract}
Large language models (LLMs) are sophisticated artificial intelligence systems that enable machines to generate human-like text with remarkable precision. 
While LLMs offer significant technological progress, their development using vast amounts of user data scraped from the web and collected from extensive user interactions poses risks of sensitive information leakage. 
Most existing surveys focus on the privacy implications of the training data but tend to overlook privacy risks from user interactions and advanced LLM capabilities. 
This paper aims to fill that gap by providing a comprehensive analysis of privacy in LLMs, 
categorizing the challenges into four main areas: 
(i) privacy issues in LLM training data, 
(ii) privacy challenges associated with user prompts, 
(iii) privacy vulnerabilities in LLM-generated outputs, 
and (iv) privacy challenges involving LLM agents. 
We evaluate the effectiveness and limitations of existing mitigation mechanisms targeting these proposed privacy challenges and identify areas for further research. 
\end{abstract}




\maketitle
\section{Introduction}


Artificial intelligence (AI) stands as a groundbreaking frontier technology, 
offering users the ability to simplify day-to-day tasks through automation and intelligent querying. 
One of the rapidly advancing AI innovations is the Large Language Model (LLM), 
which has revolutionized natural language processing, 
empowering machines to produce human-like text with exceptional precision \cite{dyde2023documentation}. 

LLMs, trained on vast datasets and characterized by extensive parameters, 
master linguistic nuances to generate sensible, coherent, and conversational responses to natural language queries \cite{yao2024survey}. 
Notable platforms like OpenAI's ChatGPT and Google's Gemini, 
boasting millions of active users, have gained immense popularity. 

Leveraging prompt engineering \cite{white2023prompt}, in-context learning capabilities \cite{brown2020language}, and retrieval-augmented generation~\cite{lewis2020retrieval}, 
LLMs have demonstrated adaptability to diverse contexts and can undertake translation, debugging, and storytelling tasks without requiring training or fine-tuning. 
While they perform impressively in various tasks, there are opportunities for improvement in addressing challenges such as hallucination, the need for domain-specific knowledge, and horizon cut-off.

\subsection{Privacy Concerns in Conventional AI/Machine Learning (ML) vs. LLMs}

Despite LLMs offering significant advancements and benefits, fairness, reliability, bias, security, and privacy concerns are becoming increasingly prominent~\cite{rahman2024survey}. 
This paper focuses on the multifaceted view of privacy in LLMs, which presents distinct challenges compared to traditional AI privacy concerns related to models and data. 
Privacy concerns in traditional AI focus on data privacy issues, such as unauthorized access to sensitive information or leakage through models via attacks like membership inference and model inversion \cite{shokri2017membership}.
In addition to these traditional risks, LLMs bring about new dimensions of privacy risks due to their advanced capabilities, 
such as a profound understanding of natural language context, human-like text generation, contextual awareness in knowledge-rich fields, and a robust ability to follow instructions \cite{yang2024give, yao2024survey}.


The utilization of LLMs in commercial applications such as OpenAI's chatGPT and Google's Gemini raises strong privacy concerns, 
particularly regarding the collection and use of personal data \cite{das2024security}. 
LLMs are often trained on extensive datasets compiled from various sources, 
including social media posts, websites, online articles, and books. 
There is a significant risk that this data, even when de-identified, can contain sensitive personal information \cite{yao2024survey}. 
Despite the sharing of such data by individuals or organizations for diverse agreed purposes, 
it is plausible that such data have been incorporated into the training data for building LLM without their explicit consent, 
constituting a privacy breach. 
Besides, the interactive nature of LLMs poses additional privacy challenges. 
As users engage with LLMs through prompts, they may inadvertently disclose sensitive information. 
The sophisticated reasoning capabilities of LLMs can infer sensitive information about users even utilizing seemingly harmless user inputs \cite{yao2024survey}.
%
%
Additionally, LLMs often integrate functionalities that involve communication with external agents or third-party modules, extending the ecosystem to perform automated tasks assigned by users.
However, such integration can further complicate privacy concerns, as each interaction has the potential to generate and disseminate sensitive data across a network of interconnected systems, each with varying levels of security and adherence to privacy concerns. 

\subsection{Motivation}

Evaluating privacy concerns in LLM systems is crucial for several compelling reasons: 
\begin{itemize}
    \item \emph{Regulatory compliance}. Regulations such as the General Data Protection Regulation (GDPR) \cite{regulation2016regulation}, the Health Insurance Portability and Accountability Act (HIPAA) \cite{act1996health}, or the EU AI Act~\cite{euai2024} impose laws on data protection and AI safety, establish substantial penalties for non-compliance and empower individuals with rights to protect their sensitive information. 
    Thus, organizations must adhere to such regulations when deploying LLM-based services. 
    \item \emph{Inherent vulnerabilities and privacy risks in LLMs}. Compared to traditional AI, LLMs introduce exacerbated vulnerabilities, 
    as they are often trained on vast amounts of potentially sensitive data. 
    This requires comprehensive analyses and methodologies to understand their strengths and weaknesses, and mitigate their vulnerabilities.
    \item \emph{Advancement of privacy technologies for LLMs}. Understanding the privacy implications of LLMs facilitates evaluating current technologies, such as differential privacy within the context of LLMs. Such insights are crucial for pinpointing gaps and driving innovation in developing robust mechanisms for privacy-preserving LLMs. 
    \item \emph{Public perception and Trust}. The widespread use of LLMs in consumer applications, particularly in critical sectors such as finance and healthcare, demands transparency and robust privacy protection to maintain public trust, as privacy violations within these domains can lead to significant financial and individual harms.
\end{itemize}

Hence, a thorough understanding of LLM systems' privacy challenges and the development of practical solutions are essential for their ethical and safe deployment.

\subsection{The Overall Objectives}

The objectives of this paper are three-fold. 
%
First, it
comprehensively maps the landscape of privacy issues associated with LLMs, analyzing these multiple dimensions: the privacy implications of the training data used, the potential for sensitive information leakage through user interactions (i.e. prompts), and privacy vulnerabilities specific to the deployment of LLMs (i.e., privacy breaches through LLM-generated output and privacy challenges involving LLM agents). 
Second, it
provides a detailed overview of the current mitigation strategies and technologies employed to address these identified privacy risks. 
%
Finally, it broadens the discussion by examining how various privacy-preserving techniques can tackle diverse privacy threats, while highlighting key research challenges and proposing avenues for future exploration.


To support these goals, we conducted a comprehensive review across various sources, including the latest research papers, tutorials, dissertations, and magazines focused on LLMs and privacy. We hope this work provides a holistic perspective for researchers, practitioners, and stakeholders engaged in the development and deployment of LLM systems.


\subsection{State-of-the-art}

The state-of-the-art studies on privacy in LLMs can be broadly categorized into privacy and security challenges in LLM models and training data.
Privacy in LLMs can be categorized further into two main groups: 
privacy risks and defense mechanisms. The former focuses
on specific privacy and security challenges within the LLM domain \cite{liu2023jailbreaking, yang2024comprehensive, carlini2021extracting,li2024personal},
while the latter focuses on 
addressing these challenges and the available defense mechanisms \cite{yao2024survey, neel2023privacy, das2024security, khowaja2023chatgpt, wu2023unveiling}.
While many surveys \cite{smith2023identifying, yao2024survey, neel2023privacy} predominantly focus on the privacy implications of LLMs and their training data, they often overlook the distinct privacy risks introduced by user interactions and LLMs' advanced capabilities \cite{carlini2021extracting}. 
Consequently, we could not identify any previous surveys or SoKs related to the privacy challenges during LLM deployment and user interaction. 


The significance of prompts cannot be overstated in LLM operation, as they facilitate the customization of pre-trained LLMs for task-specific purposes by appending a sequence of query texts\cite{lin2024promptcrypt, zhao2024explainability}. However, these prompts can often contain user-sensitive information, which can be inferred by LLM service providers, who leverage the LLMs' capabilities and access to personal data \cite{staab2024large}.
Recent advancements in the privacy domain for LLMs offer a timely and highly relevant overview of this emerging research area \cite{edemacu2024privacy}. 

Our study delves into the latest approaches and techniques in this domain, highlighting current research gaps and privacy challenges in LLMs across the categories: training data, prompts, LLM outputs, and LLM agents.

\subsection{Research Questions}

We formulated the following research questions (RQs) to address the aforementioned objectives of our study systematically.

\begin{enumerate}
    \item RQ1: How do LLMs' advanced inferring capabilities and contextual awareness impact user privacy across various dimensions, such as inference of sensitive information and memorization of user inputs? This question aims to explore the core capabilities of LLMs that may lead to privacy breaches.
    \item RQ2: What privacy challenges are associated with users actively interacting with LLMs? This question investigates how users can inadvertently expose sensitive information when interacting with LLMs. 
    \item RQ3: What state-of-the-art solutions are currently employed to mitigate the privacy risks inherent in LLMs? This question investigates the effectiveness and limitations of the existing privacy-preservation technologies for LLMs.
    \item RQ4: What are the open challenges and possible future trends to overcome these challenges? This question aims to highlight ongoing challenges and predict future technologies to preserve privacy 
    in the context of LLMs.
\end{enumerate}

\subsection{Scope}

This paper focuses on the papers published after 2022, aligning with the surge of LLMs. 
While we acknowledge the significant body of work exploring training data privacy concerns, we deliberately chose not to delve extensively into them, as they have been comprehensively covered in existing surveys \cite{smith2023identifying, yao2024survey, neel2023privacy}.
Instead, our focus lies on conducting a more in-depth analysis of other privacy challenges unique to LLMs that have not been sufficiently explored in the literature. Moreover, our study focuses exclusively on privacy issues within LLMs and does not address the broader security challenges. 
The employed methodology to identify relevant literature and establish the categories is detailed in Appendix \ref{app:source}.

\section{Overview of Privacy Challenges in LLM}
\label{sec:overview}

\begin{figure*}
    \centering
    \includegraphics[width=0.9\linewidth]{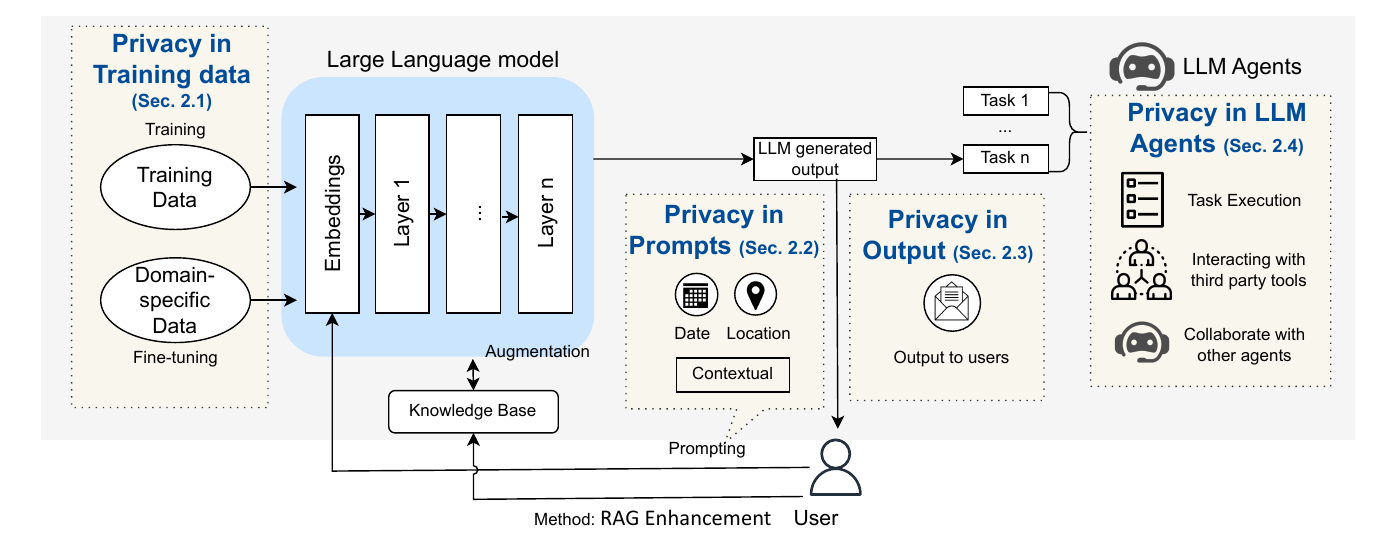}
    \vspace{-\baselineskip}
    \caption{A Multi-Faceted View of the Four Identified Privacy Challenges in LLMs}
    \label{fig:overall}
    \Description{This figure illustrates the four major dimensions of privacy challenges associated with LLMs, including risks arising from sensitive information in the training data, the potential leakage of private details through user prompts, unintended disclosures in model-generated outputs, and the broader privacy implications involved in the use of LLM-based agents in dynamic environments.}
\end{figure*}

Privacy in LLMs refers to protecting sensitive data that extends beyond traditional Personally Identifiable Information (PII) to encompass confidential, proprietary, intellectual property, and contextual or behavioral data that could reveal personal attributes or identities \cite{yan2024protecting, karamolegkou2023copyright}. This section outlines the key privacy challenges associated with LLM systems.
Our comprehensive analysis of the literature reveals four main categories of privacy challenges:
(i) privacy issues related to LLM training data \cite{neel2023privacy,feldman2020does}, 
(ii) privacy challenges in the interaction with LLM systems via user prompts in the interaction with LLM systems \cite{zhang2023s, lin2024promptcrypt}, 
(iii) privacy vulnerabilities in LLM-generated outputs \cite{priyanshu2023chatbots, yan2024protecting}, 
and (iv) privacy challenges involving LLM agents  \cite{gu2024agent, struppek2024exploring}. 
Figure~\ref{fig:overall} shows the multi-faceted view of four identified privacy challenges in LLM systems.

\subsection{Privacy in LLM Training Data}
\label{sec:priv_llm}

Privacy concerns in LLMs primarily revolve around protecting the training data utilized in their development. 
This is a key issue in commercial deployments of LLM systems, where they are trained using personal data retrieved from various sources such as social media posts and websites \cite{khowaja2023chatgpt}. 
This practice raises significant privacy concerns 
as it often involves using individuals' data without consent and contextual integrity principles when used outside their intended context \cite{nissenbaum2004privacy}.
Moreover, concerns arise regarding data storage practices, with LLM service providers such as OpenAI storing personal information potentially conflicting with GDPR \cite{avastChatGPTsPeoples}. 
Besides the privacy concern arising from the utilization of public data and the data owner's consent, LLM models are susceptible to significant privacy vulnerabilities, such as memorization and the inadvertent leakage of PII or confidential data \cite{das2024security}. 

\subsection{Privacy in Prompts}

There are potential privacy leakages from user-provided input prompts during interactions with LLM systems \cite{staab2023beyond}. The ever-increasing inference capabilities of LLM systems and their widespread adoption across various domains have raised significant privacy concerns through user prompts.
User prompts are often exposed to the service providers, raising questions about whether current LLM systems could breach users’ privacy by inferring personal attributes from the prompts. 
Despite data transmission and storage encryption, LLM service providers retain the knowledge of the actual data contents, undermining individuals' or entities' trust in these services. 

The interactive utilization of LLM systems introduces a new set of privacy risks during inference time. 
LLM systems are fed with diverse types of information from various sources in their prompts, potentially revealing more contextual data beyond the direct sensitive data in the prompts. Recent findings in \cite{staab2023beyond} have demonstrated that modern LLM systems can be leveraged for highly accurate predictions of personal attributes, even from seemingly innocuous data. 
Previously, human involvement was necessary for inferring private attributes. 
However, these models are now swift and economical enough to automate such inferences at scale \cite{staab2024large}. 
Simultaneously, users remain unaware of these concerns, inadvertently sharing texts containing information easily inferable by LLM systems.
For evidence of privacy vulnerability via prompts, The New York Times\footnote{https://www.nytimes.com/2023/03/31/technology/chatgpt-italy-ban.html} reported instances where personal information, including chatbot conversations and login credentials, was unintentionally exposed. 

Given these challenges, we believe the research community is responsible for fostering a paradigm shift in LLM-centered privacy research. 
New research should extend beyond examining privacy risks associated only with models and data 
to explore privacy vulnerabilities inherent in other aspects (e.g., user interactions) of LLMs.
Prompts are one significant aspect as they directly reveal substantial information. These approaches should prioritize user-friendliness, ensuring users can easily and quickly verify whether they disclose private information in their prompts.

\subsection{Privacy in LLM-generated Outputs}
\label{sec:over_llm_output}

Within the scope of privacy concerns regarding LLM-generated outputs, we delve into issues such as the retention, extraction, and retrieval of sensitive user data in outputs (i.e., memorizing sensitive information from user prompts) and the inadvertent inclusion of sensitive information in the LLM outputs.

We consider LLM-generated output privacy as a distinct aspect of LLM privacy due to the following reasons: 
1) Users may enter sensitive information in prompts without expecting it to appear in the output. 
However, LLMs may still include such information in their responses. 
This issue is prevalent in In-context Learning (ICL), where users' private data in prompts help adapt the model to specific tasks with only black-box access.  
Due to the context-specific examples in the prompts, the output can include labels of sensitive data samples from the prompt data for ICL.
2) In commercialized LLM settings like the GPT store\footnote{https://openai.com/index/introducing-the-gpt-store/}, specialized LLM models are fine-tuned using private data through methods like ICL, fine-tuning, and Retrieval-Augmented Generation (RAG). 
These techniques integrate user queries with demonstrations or relevant documents from a knowledge base to enhance LLM-generated responses. 
However, this customization poses a risk of exposing private information from a smaller user base to a broader audience. Although LLM products should ideally avoid generating harmful outputs such as sensitive information, legal, medical / health, financial advice, and misinformation, they can be customized to produce such output.
3) Even when users employ techniques to protect sensitive information in input prompts, LLM-generated outputs remain vulnerable to
exposure to malicious service providers, third-party tools, external parties, or hackers\cite{yao2024privacy}. Service providers can potentially disclose sensitive information in output prompts by analyzing training data or accessing external sources, even with partially sensitive data from prompts. 
For example, using protection mechanisms at the user end, users can restrict LLM systems from retaining sensitive data 
(e.g., passport identifier)
in input prompts. However, LLM systems may still inadvertently reveal sensitive information such as the birth year or age in the output due to its processing \cite{wu2023unveiling}.

\subsection{Privacy in LLM Agents}

\begin{figure}[!ht]
\vspace{-\baselineskip}    
    \centering
    \includegraphics[width=1\linewidth]{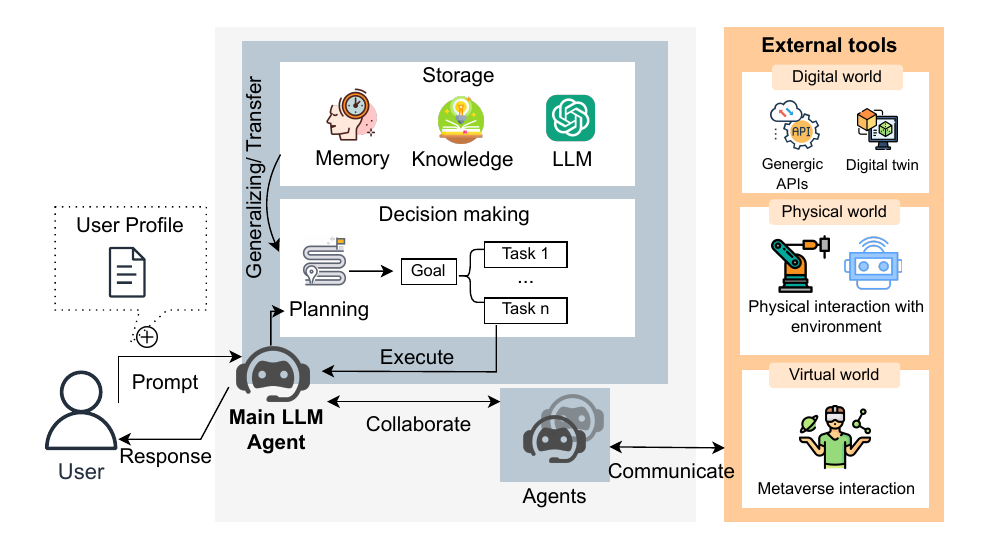}
    \caption{The Workflow of LLM Agents}
    \label{fig:agent}
    \Description{This figure depicts the typical workflow of LLM-based agents, outlining how they interpret user inputs, interact with tools or external environments, process intermediate steps through reasoning or planning modules, and ultimately generate responses or perform tasks based on the given objectives.}
\end{figure}

Recent advances in LLM Systems have led to the development of agent-based solutions such as WebGPT \cite{nakano2021webgpt}, AutoGPT \cite{githubGravitasAutoGPT}, and GPT Plugins (e.g., WebPilot \cite{zhang2024webpilot}). 
As illustrated in Figure \ref{fig:agent}, in these systems, a main LLM agent divides a user prompt into 
individual tasks and transmits them to secondary agents (which can also be LLM-based). 
These agents then
leverage a range of powerful tools, including other agents and third-party applications (e.g., mobile apps, 
web browsers, sensors, metaverse interfaces, code interpreters, and API plugins) to interact with external
environments and carry out real-world tasks assigned by the main LLM agent.

The transition from text-based interactions to those involving sensitive data with agents operating in the real world raises significant privacy concerns \cite{ruan2023identifying}.
For example, physical interaction-based LLM agents heighten privacy risks by collecting real-world data such as facial photos (cameras) and speech audio (microphones) 
\cite{stores1}. 
A significant challenge in such an agent-based system is the risk of unauthorized propagation of sensitive data across different components of the system~\cite{li2024personal}. 
Indeed, agents autonomously interact with external tools, different data modalities, other AI models, and APIs, making tracking and controlling data flow difficult.
Users may provide private inputs through prompts, and when processing their requests, an LLM agent may share their data with external tools or across interconnected systems without their explicit awareness. 
The agent's ability to autonomously reason, execute multi-step actions, and retrieve information from various sources further amplifies the privacy risks, as data may persist across interactions in ways not immediately apparent to users.
LLM agents pose risks of data persistence, where sensitive information from past interactions may resurface in future prompts, especially when contextual memory is retained across sessions or users \cite{li2024personal}. 

Hence, LLM agents extend beyond traditional LLM output privacy concerns by persisting memory, unintended exposure across sessions and contexts, autonomously interacting with external tools, and operating in real-world physical interactions with environments \cite{stores1}. Therefore, stricter and more diverse mitigation strategies are needed to address privacy concerns in LLM agents.

\section{Privacy Challenges in LLM and Mitigation}

This section discusses privacy challenges under four broad categories and reviews existing mitigation techniques in the literature.

\subsection{Privacy Issues in LLM Training Data and Mitigation}
\label{sec:llm_data}

\begin{figure}[!ht]
    \vspace{-\baselineskip}
    \centering
    \includegraphics[width=0.9\linewidth]{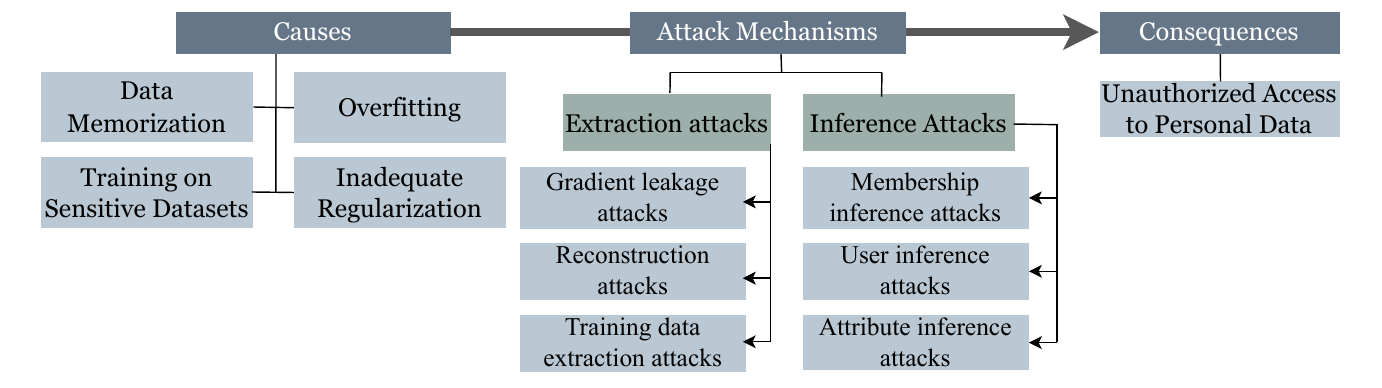}
    \caption{The Privacy Challenges in LLM Training Data}
    \label{fig:llm_challenges}
    \Description{This figure explores the privacy challenges associated with LLM training data by examining their root causes, the potential attack mechanisms that exploit these vulnerabilities, and the resulting consequences, offering a structured view of how privacy risks emerge and impact model behavior.}
    \vspace{-\baselineskip}    
\end{figure}

We analyze privacy issues in LLM training data by distinguishing between the causes, 
attack mechanisms, and consequences of privacy vulnerabilities. Data memorization is one of the primary \textit{causes} of privacy risks, where models inadvertently retain and reproduce sensitive information from training data. Privacy attacks act as \textit{mechanisms} that adversaries use to exploit LLMs to extract sensitive information. 
%
%
These attacks can lead to \textit{consequences} such as privacy leakage. For example, unauthorized access to personal or confidential data can compromise privacy.
Figure \ref{fig:llm_challenges} illustrates these interrelated aspects of privacy risks in LLMs. 
Although prior research works \cite{neel2023privacy, das2024security, smith2023identifying} extensively discuss privacy threats and mitigation, they have not examined this causal structure. This section summarises key insights, emphasizing how data memorization contributes to privacy vulnerabilities and how various attack techniques facilitate privacy leakages.

\subsubsection{Potential Causes of Privacy Issues}

Many privacy issues in LLMs arise due to data memorization, which can be exploited by privacy attacks, ultimately resulting in privacy leakages \cite{kassempreserving}. 
Data memorization is an intrinsic characteristic of many ML models, as the training process involves retaining specific information from input data to make accurate predictions \cite{momorizationdl}. 
However, while users expect LLMs to generate novel content that is semantically similar to the training data, the verbatim memorization and reproduction of learned phrases can potentially expose sensitive data and breach privacy \cite{neel2023privacy,feldman2020does}. Hence, memorization is particularly problematic when LLMs are trained on personal information datasets.

Beyond data memorization, privacy leakage can stem from training on sensitive datasets without proper sanitization \cite{carlini2021extracting}. Overfitting and inadequate regularization (e.g., dropout, weight decay) further exacerbate the risk by making models more likely to retain and expose specific training data \cite{kaya2020membershipinference, tan2022parametersprivacyprovabletradeoff}.

\paragraph{\textbf{Mitigation}}
\label{miti:potential_leak}

A fundamental approach to mitigating privacy issues in LLM training data is to reduce data memorization, which is the basis for many attacks. We first examine techniques designed to address the \textit{challenge of data memorization}.
One straightforward approach is to deduplicate training datasets, as data redundancy can exacerbate model memorization. Even a 10-fold data duplication can lead to a 1000-fold increase in memorization \cite{kandpal2022deduplicating}. Another method is early detection of memorization during the training phase, enabling practitioners to take corrective actions such as discarding memorized points, reverting to a checkpoint, or halting the training process for adjustments \cite{neel2023privacy}. 
The utilization of filtering techniques such as the bloom filter method (which is a probabilistic data structure used to detect and filter out memorized or sensitive data efficiently) \cite{ippolito2022preventing} is another defense mechanism for memorization, 
scanning the training datasets to check if the model's next token forms an $n$-gram in the training set and selecting an alternate token by sampling from the model's posterior if it does. 
\textit{However, this method does not guarantee privacy and can be bypassed with plausible, minimally modified prompts, failing to prevent training data leakage entirely.} Another promising approach for mitigating memorization challenges is differential privacy (DP: detailed in Table \ref{tab:glossary_table}), which adds noise to data during training, providing mathematical guarantees for privacy but reducing utility \cite{zhao2022provably}. LLM editing offers yet another approach, in which neurons corresponding to memorization and storage of specific training data knowledge can be directly edited (i.e., by altering internal parameters) \cite{chang2023localization}.

A key consequence of data memorization is personal or confidential data leakage, which poses a significant privacy risk. 
While general techniques such as deduplication \cite{kandpal2022deduplicating, lee2021deduplicating} can help mitigate model memorization and reduce the risk of personal data leakage, some studies specifically target this issue by employing advanced approaches, such as fuzzy logic deduplication \cite{brown2020language} to prevent the retention of sensitive information.
Other methods to prevent privacy leakage include data cleaning, PII scrubbing, and filtering with restrictive terms of use \cite{yan2024protecting, asthana2025adaptive}. Data cleaning enhances privacy by correcting errors, implementing anonymization, and following secure practices to protect sensitive information \cite{yan2024protecting, pilan2022text}. PII scrubbing filters \cite{sentryDataScrubbing, asthana2025adaptive, subramani2023detecting} use Named Entity Recognition (NER) to tag and remove PII. However, these tools cannot guarantee complete removal, and AI advancements can infer PII from non-PII data. Therefore, minimizing data retention and enforcing purging policies are critical to reducing the risk of breaches and unauthorized access.
\textit{While most of these approaches are applied to smaller text corpora, the use of extensive training data in LLMs presents challenges. More empirical studies are needed to evaluate their effectiveness in such large-scale models with larger datasets.}

Recently, knowledge unlearning techniques \cite{eldan2023whos, zhang2023composing, yao2023large} have been utilized in LLMs to force models to forget specific knowledge without requiring full retraining. These knowledge unlearning techniques involve randomly sampling data from the training corpus and performing gradient ascent, altering the direction during language modeling on target token sequences. However, the success of this method depends heavily on specific target data and the domain of the data to be unlearned \cite{smith2023identifying}.

Another recent method for addressing PII leakage in LLMs is ProPILE \cite{kim2024propile}, a tool that allows data subjects to assess the inclusion and potential leakage of personal data in LLM systems during deployment. ProPILE considers the linkability and structurability of data in inferring PII. Singh et al. \cite{singh2024whispered} proposed a multi-faceted whispered tuning approach by integrating PII redaction, DP, output filtering, and architectural improvements to enhance privacy preservation. Likewise, OpenAI utilizes filtering and fuzzy deduplication techniques to remove PII from the corpora utilized for model training \cite{brown2020language}.
\textit{Despite these efforts, achieving complete prevention of privacy leaks remains challenging. 
Hence, in-depth studies are necessary to design effective defense techniques for PII leakages on LLMs while maximizing user education and involvement.}

\subsubsection{Potential Privacy Attacks}

To exploit data memorization in LLM training data, various privacy 
\textit{attack mechanisms} have been developed, each employing different techniques and targeting specific vulnerabilities.
While LLMs utilize Deep Learning (DL) for model building, all traditional attacks on DL models may not directly apply to LLMs due to limited access to model parameters and the prevalent use of application programming interfaces (APIs) in most LLMs. However, certain LLM service providers offer open-source models (e.g., LLaMA, BERT, DeepSeek, Mistral 7B
), making DL-related attacks extended to the LLMs. 
Privacy attacks on LLM can be broadly categorized into two classes: extraction attacks and inference attacks.  

\paragraph{\textbf{Extraction Attacks}}

Extraction attacks involve adversaries attempting to extract sensitive information or insights from the LLM model by accessing model gradients, the training data, or adversarial prompting. LLMs may inadvertently capture and reproduce sensitive information in the training data, potentially raising privacy concerns during the text generation process \cite{neel2023privacy}. Various types of data extraction attacks include gradient leakage attacks \cite{deng2021tag, zhu2019deep, balunovic2022lamp}, reconstruction attacks \cite{chu2024conversation, liu2024making, balle2022reconstructing}, and training data extraction attacks \cite{carlini2021extracting, zhang2023ethicist, yang2023code, nasr2023scalable}. Gradient leakage attacks exploit DL optimization algorithms, accessing gradients to leak sensitive data. However, this attack is challenging when the model architecture is unknown (i.e., closed-source) \cite{deng2021tag}. Reconstruction attacks aim to recover private training data by analyzing model parameters, gradients, or generated output. Even without direct access to LLMs, attackers can use shadow modeling with external datasets and feedback from the target LLM \cite{zhang2022text}. Training data extraction attacks target sensitive information in training data by strategically querying LLM systems via prompts to retrieve specific examples, including personal information \cite{das2024security, carlini2021extracting}. Jailbreaking \cite{wei2024jailbroken} and prompt injection \cite{shin2020autoprompt} security attacks make LLMs more vulnerable to extraction attacks and introduce substantial privacy risks with broad and potentially severe implications even without direct model access.
%
Jailbreaking attacks exploit both direct and indirect techniques to bypass the safety and alignment constraints of LLMs. These attacks often use manually or automatically crafted adversarial prompt templates designed to deceive the model into generating harmful content or revealing sensitive information, potentially including memorized data from its training corpus \cite{liu2023jailbreaking}.
%
In terms of extracting private information, a multi-step jailbreaking approach has bypassed ChatGPT’s safety mechanisms, successfully extracting PII from ChatGPT and LLM-powered search engines \cite{li2023multi}. Another study tested 3,700 templated jailbreak prompts, analyzing their effectiveness across different LLMs \cite{rao2023tricking}. Additionally, an automated framework, ReNeLLM, was proposed to generate effective jailbreak prompts using prompt rewriting and scenario nesting to infer private information \cite{ding2023wolf}.

\paragraph{\textbf{Inference Attacks}}

Inference attacks aim to acquire knowledge or insights about a model or data characteristics by observing the model's responses or behavior \cite{yao2024survey}. Common inference attacks in LLMs include membership inference attacks (MIA), attribute inference attacks, and user inference attacks \cite{wang2024pandora, jagannatha2021membership, mireshghallah2022empirical}. 

In MIAs, an adversary attempts to predict whether a particular data point (e.g., a particular sentence or document) is a member of a target model’s training dataset \cite{duan2024membership} (Detailed in Table \ref{tab:glossary_table}).
MIA can be conducted solely based on the text generated by LLM systems. While MIAs generally perform no better than random guessing for LLM training data due to the vast datasets used \cite{duan2024membership}, their effectiveness improves when small datasets are used for fine-tuning for specific tasks, known as `User Inference Attacks' \cite{kandpal2023user}. These attacks are prevalent in fine-tuned LLMs using techniques such as full fine-tuning, RAG, and ICL techniques \cite{fu2023practical, duan2023privacy}.

User inference attacks present a significant privacy risk for \textit{LLMs fine-tuned on a smaller group of user data}, where an attacker tries to determine if a specific user participated in the fine-tuning process. Minimal user samples and black-box access to the fine-tuned model are sufficient for this attack to succeed \cite{kandpal2023user}. This poses a privacy risk if the fine-tuning task reveals sensitive information about users, such as a model fine-tuned exclusively on users with a rare disease.

Attribute inference attacks (Detailed in Table \ref{tab:glossary_table}) pose another privacy risk for LLM systems, as adversaries can use these attacks to identify users' attributes through prompts using publicly available partial data of users \cite{staab2023beyond}. Robin et al. \cite{staab2023beyond} demonstrated
that personal details can be inferred accurately from current LLMs.
While prompt injection is often discussed in the literature as a backdoor security attack, it can also be used for attribute inference.


\paragraph{\textbf{Mitigation}}

Differential Privacy (DP) techniques have become a popular approach in addressing such privacy attacks\cite{zanella2020analyzing, singh2024whispered, shi2021selective, li2021large}.
Besides DP, techniques to reduce memorization (discussed in Section \ref{miti:potential_leak}) can also help prevent these privacy attacks. For example, Jagielski et al. \cite{jagielski2022measuring} discovered that using larger datasets during fine-tuning or more extended training with non-sensitive data could effectively mitigate privacy concerns by reducing the MIA. 
Furthermore, test-time defense and instruction processing are also crucial for mitigating privacy attacks \cite{yao2024survey}. Test-time defenses filter malicious inputs, detect abnormal queries, and post-process LLM outputs. Instruction pre-processing transforms user prompts to eliminate adversarial contexts or malicious intents \cite{robey2023smoothllm, li2022text}.
Enhancing LLM architectures can also improve safety by integrating LLMs with knowledge graphs \cite{zafar2023building} and cognitive architectures \cite{romero2023synergistic}. Furthermore, the LLM training process can employ robust optimization methods such as adversarial training \cite{yoo2021towards} and robust fine-tuning \cite{dong2021should} to prevent malicious text attacks.

\textit{However, these techniques have limitations, such as challenges in effectively handling adversarial inputs, ensuring scalability in large models, and maintaining a balance between privacy and model performance, particularly in dynamic, real-world scenarios.}

Federated learning (FL) may potentially solve privacy attacks in LLM systems by shifting processing from central servers to users \cite{ro2024efficient, vu2024analysis}. However, its application to LLMs faces challenges due to inadequate FL framework support, handling vast data and complex models, and optimizing communication and computational resources. Some studies \cite{fan2023fate, ye2024openfedllm} have explored FL for LLMs, but its viability remains uncertain. Federated LLM fine-tuning may enable parameter-efficient fine-tuning with limited resources \cite{kuang2023federatedscope}.

Fully Homomorphic Encryption (FHE) has been explored as a privacy-preserving solution \cite{app132413146, rho2024encryption, rovida2024transformer} for language models such as DistilBERT \cite{sanh2019distilbert} (66M parameters) and BERT (110M parameters) \cite{devlin2019bertpretrainingdeepbidirectional}. While FHE has been applied to smaller models with limited datasets \cite{app132413146, rho2024encryption, rovida2024transformer}, scaling it to larger models, such as GPT with massive datasets, remains challenging due to the computational complexity. Outsourcing computation can also be securely conducted for small language models (LM): for instance, an encrypted RNN text classifier was shown to predict on homomorphically encrypted input without accuracy loss \cite{li2025privacy}, and such overhead remains manageable at this scale.
With a compact model, full on-device inference becomes realistic, retaining user data locally and mitigating cloud-related vulnerabilities \cite{xu2023llmcad,llmmobile,yi2025edgemoe}. Conversely, LLMs with billions and trillions of parameters often demand cloud-based infrastructures and massive datasets, introducing more complex privacy concerns.

Mitigation strategies for jailbreaking attacks are extensively studied in the literature and can be categorized into prompt-level and model-level approaches, depending on whether they modify the protected LLM. Prompt-level defenses (e.g., input sanitization) \cite{defending} focus on filtering adversarial prompts using rule-based detection or classifier-based approaches.  
Model-level defenses (e.g., adversarial training) \cite{jiang2025robustkv} involve enhancing LLMs through safety training, reinforcement learning with human feedback (RLHF), and adversarial fine-tuning to improve robustness against attacks. Constitutional AI frameworks \cite{bai2022constitutional} and automated moderation systems also help enforce ethical guidelines and restrict unintended model behaviors. To protect private information from jailbreaking, prior work has explored mostly prompt-level mitigation: defending with additional prompts and using a harmfulness classifier to filter malicious inputs \cite{rao2023tricking, ding2023wolf}. The study in \cite{li2023multi} outlined potential approaches at both model-level defenses and prompt intention detection mechanisms to prevent the disclosure of private information.

\begin{tcolorbox}[colback=gray!15!white,colframe=gray!75!black,title=Key Takeaways]
LLM systems have shown an ability to memorize training data, raising concerns about the unintentional disclosure of sensitive information and privacy breaches. 
Despite the efforts to address these issues with methods such as differential privacy, knowledge unlearning, and test-time defenses, adversaries may still infer private data through adversarial prompting. 
Researchers should focus on establishing responsible practices that safeguard privacy while preserving the utility of LLMs trained on large-scale datasets.

\end{tcolorbox}

\subsection{Privacy Challenges in the Interaction with LLM systems via Prompts and Mitigation}
\label{sec:llm_prompt}

This section focuses on privacy challenges when users interact with pre-built LLM systems via prompts. We identify three types of privacy challenges associated with prompts in the literature: the direct leakage of sensitive data in prompts, the potential inference of sensitive information, and the leakage of contextual information from user devices and activity logs.

\subsubsection{The Direct Leakage of Sensitive Data in Prompts}

Users might inadvertently reveal personal data through prompts when interacting with the system, often due to a lack of awareness or reluctance to undertake additional efforts to minimize personal information in prompts. Even individuals with strong technical backgrounds, often in IT industries, reportedly leaked sensitive company data through ChatGPT prompts in a late 2022 incident\footnote{https://tinyurl.com/yny4mjdh}.

This challenge became more prominent with the integration of LLMs into interactive computing systems such as conversational agents, making them widely accessible for everyday tasks. Users tend to disclose more private information when engaging with LLM-based conversational agents due to the tools' high utility and human-like interactions \cite{zhang2023s}. Moreover, many users are unconcerned about revealing personal data in prompts, maybe under the misconception that LLM companies no longer collect interaction data, based on various published statements~\cite{wu2023unveiling}.
However, despite the prevailing belief, often the company's privacy policies indicate that user prompts are periodically integrated into the training process by default \cite{wu2023unveiling, zhang2023s}. 
The extent of sensitive data revealed in user prompts is evidenced in the study by Zhang et al. \cite{zhang2023s}. 
To identify PII leakage in prompts, they analyzed real-world ChatGPT conversations. 
The study uncovered significant amounts of sensitive and personal data in LLM prompts, providing a comprehensive overview of user disclosure behaviors. 

Protecting personal data in prompts is imperative. Conversations between users and LLMs can become embedded in LLM parameters during training, making them vulnerable to adversarial attacks. Hence, data protection measures should be implemented at the user end to protect data from service providers and external parties.

\paragraph{\textbf{Mitigation}}

A straightforward mitigation technique for potential direct leakage of sensitive data is input validation and sanitization. Some approaches rely on LLM service providers to implement these solutions, so their efficacy depends on the trustworthiness of the providers or the LLM itself. Other methods acknowledge that privacy vulnerabilities can arise from both LLM service providers and external parties, offering a safer solution than those relying on the trustworthiness of the service providers. One sanitization approach involves using NER or predefined policies and rules to identify and remove specific sensitive details in prompts. Research works \cite{lin2024promptcrypt, chen2023hide, li2024human} and commercial products \cite{stracSecureSensitive, lakeraIntroducingLakera} focus on prompt privacy using this method.

To address the dilemma of potentially compromising prompt utility by removing all personal information from user prompts, EmojiCrypt \cite{lin2024promptcrypt} converts sensitive information in prompts into an encrypted format using emojis and mathematical operations. This method retains informative value for LLMs while making the data incomprehensible to humans. However, it relies on trusting the LLM for encryption, mainly aiming to conceal information from other users. Other constraints include a limited symbolic vocabulary and the risk of introducing inaccuracies.

Another lightweight anonymization technique proposed in \cite{chen2023hide} aims to protect prompt privacy through substitution or masking. 
The approach involves two core techniques: hiding private entities in prompts using generative schemes and NER models and seeking private entities for de-anonymization. Both models are implemented on the user's system. After anonymizing the text with the Hide model, the result is sent to the LLM, and the Seek model de-anonymizes the output. This approach commendably does not trust the LLM. \textit{However, it relies on a small corpus or NER for identifying sensitive data, which limits its ability to conceal other words and sentence structures. Additionally, replacing entity words in prompts can challenge tasks that rely on precise semantics.}

Another potential approach involves designing secure prompt templates and sensitive data redaction tools to minimize personal data leakage \cite{stracSecureSensitive}. This could include using structured data for inputs instead of free text. Variables can be safely inserted into the prompt template without giving users direct control over the prompt structure, allowing for easy removal of sensitive data.

Another technique is a combination of small local LMs and remote LLMs.  \citet{hartmann2024can} proposed a privacy-preserving mechanism where a local small LM at the user's end personalizes outputs without disclosing confidential information to the remote model. 
They explored three algorithms: (1) generating a high-level description of the query locally and using the remote LLM for few-shot examples, (2) creating a similar novel problem with new unlabeled examples for the remote LLM, and (3) maintaining the query’s structure while substituting private information with placeholders.
\textit{These methods may be domain-specific, rule, or NER-dependent, requiring users to have sufficient knowledge to manage them.}

TextObfuscator \cite{zhou2023textobfuscator} protects user privacy using text obfuscation techniques. It learns private representations that obscure original words while retaining their functionality. This is achieved by identifying prototypes for each word and clustering functionally similar words around the same prototype. Random perturbations are then applied to these clusters, obscuring original words and maintaining functionality. \textit{However, this approach requires significant computational power on local machines, only protects word privacy during inference.}


Recently, \citet{ruoyan2025practical} integrated FHE and provable security theory with parameter-efficient fine-tuning to propose a secure inference scheme for LLMs, ensuring the protection of both user-side inputs and server-side private parameters. 
Likewise, \cite{zhang2024secpe} designs efficient FHE counterparts for the
core algorithmic building blocks of prompt ensembling. In \cite{chen2022x, hao2022iron, de2024privacy}, the authors explored privacy-preserving inference using transformer models, incorporating FHE and secure multi-party computation to protect the input text in user prompts. \textit{However, computing and communication efficiency remain a challenge when handling multiple prompts.}



\subsubsection{The Potential Inference of Sensitive Information}

Aside from direct leaks of sensitive data, the capabilities of LLMs enable even seemingly insignificant data to unveil sensitive information through inference \cite{kroger2022personal}. Leveraging their advanced capabilities, LLMs can infer various personal attributes from prompts with the help of extensive sets of unstructured training data. For instance, from the prompt, ``I always get stuck there waiting for a hook turn,'' LLMs could infer that the individual is in Melbourne, as hook turns are a distinct traffic maneuver primarily employed there \cite{li2024human}. 
This inference ability, combined with widespread LLM availability, reduces the costs of private data inferences, enabling adversaries to scale beyond limitations imposed by costly human profilers. Existing rule-based or NER-based redaction tools often prove inadequate in protecting users from such sensitive inference, as they struggle to detect revealing, context-dependent cues.

\paragraph{\textbf{Mitigation}}

Most existing works utilize LLMs themselves as a mitigation technique to identify sensitive information that can be inferred from prompts.
A study in \cite{staab2023beyond} shows the ability of pre-trained LLMs to infer personal attributes from the text given at inference time. Their extensive experiments demonstrated LLMs' capacity to infer personal attributes from real-world data, even when the text is anonymized using commercial tools. 
While this paper highlighted the potential for sensitive data leakage from prompts, it did not delve into mitigation techniques.

Building on the work in \cite{staab2023beyond}, a later study \cite{staab2024large} introduced a framework for anonymizing texts using an adversarial feedback-guided approach. This method leverages LLMs' strong attribute inference abilities to guide a separate anonymizer LLM. The process involves two steps: an `LLM adversary' performs private attribute inference, and an `anonymizing LLM' adjusts the text to obscure or generalize the inferred cues. \textit{However, this method doesn't fully address where the LLM adversary should be located (user or service provider) or the utility of the anonymized text. Using smaller, fine-tuned LLMs locally could improve this process by identifying inference data before it reaches the remote LLM, and incorporating LLM bias and prior knowledge may enhance inference detection.}

\subsubsection{The leakage of contextual information}

LLMs often rely on contextual information to generate personalized and impactful content, leveraging their in-context learning capabilities. The contextual information may include users' personal data, such as app usage data (e.g., shopping lists, logs, and calendar events). Most users include contextual data in prompts for improved performance, as it is more parameter and data-efficient than fine-tuning \cite{duan2023privacy}. 

Given the granular nature of contextual data, privacy risks naturally arise. It's essential to understand these risks, mainly because the data used in prompts often comes from smaller, private datasets, unlike the large public corpora used for pretraining LLMs. This small set of private data is more vulnerable to inference attacks. 
Therefore, striking a balance between performance and privacy is crucial.

\paragraph{\textbf{Mitigation}}

In the literature, techniques for protecting contextual data in prompts include hashing operations \cite{yim2023privacy}, prompt ensembling \cite{duan2023privacy}, and employing a personalized LM on the local with contextual information to improve the remote LLM responses\cite{zhang2024cogenesis}.

Yim et al. \cite{yim2023privacy} used a hash operation on context data before sending it to the LLM, ensuring only hashed values are transmitted. The LLM's response also includes hashed values that are then reverted to their original form. This method protects personal data privacy while enabling personalized responses. 
Duan et al. \cite{duan2023privacy} employed a prompt ensembling method to mitigate MIA on prompted data with contextual information. This method reduces the MIA success rate to near-random guessing levels. The prompt ensembling technique aggregates prediction probability vectors over multiple independent prompted models into an ensemble prediction. 
In CoGenesis \cite{zhang2024cogenesis}, smaller, personalized LMs on user devices access private data and activity logs, while advanced general LLMs operate in the cloud, receiving only general instructions and providing high-level knowledge. This allows for collaborative content generation, where the user-end model uses context and responses from the LLMs to produce personalized outputs. 

\textit{Existing techniques face key challenges. Hashing methods \cite{yim2023privacy} rely on trusting the LLM and only protect against eavesdropping. Prompt ensembling \cite{duan2023privacy} lacks formal privacy guarantees and adds computational overhead. CoGenesis \cite{zhang2024cogenesis} depends on logit access, limiting its use with closed-source LLMs, and tested on synthetic data, raising questions about real-world applicability. 
}

\begin{tcolorbox}[colback=gray!15!white,colframe=gray!75!black,title=Key Takeaways]
User prompts submitted to LLM systems may contain sensitive information. 
Given the advanced inference capabilities of modern LLMs, 
even seemingly benign or contextual data can lead to the disclosure of private details,
making traditional NER or rule-based detection methods inadequate. 
As mitigation, researchers are investigating the use of small, fine-tuned LLMs with ICL/RAG techniques at the user's end to assess the risk of sensitive information inference and enhance prompt privacy.
\end{tcolorbox}

\subsection{Privacy Vulnerabilities in LLM-generated Outputs and Mitigation}
\label{sec:llm_decision}

Privacy vulnerabilities in LLM outputs arise when sensitive information, whether learned from training data or newly produced in response, is inadvertently disclosed to unintended recipients. This can happen if the model memorizes personal or confidential details in a generated text or synthesizes proprietary content (such as internal strategies, algorithms, or trade secrets) accessible beyond authorized channels. In either case, allowing these outputs to be publicly visible or logged by external systems exposes private data to potential misuse.

\subsubsection{Revealing Sensitive Information in LLM-generated Output}

As reported in many studies \cite{priyanshu2023chatbots, li2024human, zhang2023s}, LLMs can memorize sensitive data from prompts and include them in their output despite users explicitly requesting them not to memorize their data in prompts.
Priyanshu et al. \cite{priyanshu2023chatbots} evaluated chatbot responses' compliance with privacy regulations, particularly when user data are used as few-shot samples for ICL. They found that ChatGPT reproduces PII accurately 57.4\% of the time, decreasing to 30.5\% with regulation prompts and 15.2\% with explicit removal prompts. 
This behavior can be considered a privacy violation as it occurs without user consent. Moreover, there's the potential for additional privacy breaches if these outputs are compromised through attacks, eavesdropping, or retraining the LLM model with user data.

This risk is heightened when the model is fine-tuned with private data for domain-specific applications, such as when organizations use LLMs for specialized internal tasks. Even within a controlled setting, sensitive data can be exposed to all users with access permissions. For example, an LLM-driven HR chatbot fine-tuned with employee details could inadvertently allow employees to view sensitive information about their peers' salaries and benefits\footnote{https://medium.com/snowflake/handling-sensitive-data-with-llms-aa765f8ce840}. Without proper controls to restrict access, the model could expose this information to all users.

This risk increases when LLM-generated outputs are shared with third-party services, especially in integrated systems, such as chat platforms or APIs. Unlike LLM agents, which autonomously interact with external tools and the environment, LLM output privacy concerns stem from the uncontrolled dissemination of generated responses to third-party services without the user’s explicit awareness, increasing the risk of unintentional data leakage.
Ensuring control and transparency in accessing privacy data with user consent is crucial, especially since the output can potentially be sent to numerous third-party applications once it leaves the LLM. Hence, it is essential to implement techniques to manage the data workflow and keep users informed about it.

\paragraph{\textbf{Mitigation}}

Mitigation techniques for privacy challenges in LLM systems' outputs are still nascent, with only a few research studies addressing this issue. Given the lack of user control over the black box settings in LLM services, controlling the LLM systems' decisions is challenging. The only feasible method for users to influence the model output is solely through prompts. Otherwise, service providers are responsible for managing the output. 

Yao et al. \cite{yao2024privacy} proposed an initial approach to protect LLM decisions in a black-box manner. They define decision privacy and investigate instance obfuscation strategies for decision privacy. They tackled the privacy concerns of decision-making by appending an `obfuscator' (another random text) to the original prompt. 
The method uses obfuscators to alter LLM decision distribution, preventing adversaries from inferring the correct decision while the data owner resolves it from obfuscated inputs.

Likewise, to protect the privacy of decisions made in ICL, the studies in \cite{wu2023privacy, tang2023privacy} concentrated on differentially private aggregation methods to prevent the direct extraction of private data. Wu et al. \cite{wu2023privacy} introduce the DP-ICL framework, which can aggregate and release responses without heavily relying on any individual examples provided in the demonstrated private data. The fundamental concept behind the DP-ICL framework involves producing differentially private responses by aggregating noisy consensus from an ensemble of LLMs' outputs, each operating on separate sets of examples. 

\textit{
In summary, current mitigation techniques for protecting private information in LLM outputs illustrate a clear trade-off between privacy and utility.
Specifically, the approach in \cite{yao2024privacy}, designed mainly for static tasks (e.g., text classification), does not adequately address the generative nature of LLM outputs, risking exposure of confidential data through dynamic outputs.
Moreover, cross-session data retention poses an additional threat to data privacy: if previous conversation states, memory buffers, or hidden user context are not carefully managed or sanitized, sensitive information can reappear in later responses, bypassing any initial privacy safeguards. Consequently, while these methods offer valuable insights, they require further refinement to incorporate robust safeguards against unintended data exposure.}

\begin{tcolorbox}[colback=gray!15!white,colframe=gray!75!black,title=Key Takeaways]
Recent techniques such as ICL and RAG have exacerbated privacy concerns with LLM-generated outputs, as users increasingly provide sensitive data for domain-specific tasks. 
The opaque nature of black-box LLM services limits user control, making it difficult to manage or audit model-generated decisions externally. 
To mitigate these risks, 
users may obfuscate their input before submission.
Additionally, LLM providers may consider implementing techniques like DP to reduce potential leakage in outputs.
\end{tcolorbox}

\subsection{Privacy Challenges while Involving Agents in the LLM System Tasks and Mitigation}
\label{sec:llm_agents}

This section addresses privacy challenges that arise when users interact with LLM agents to accomplish tasks that communicate with the external world. 
These privacy challenges in LLM agents can be categorized into three areas: automated task execution via LLM agents, adversarial interactions of agents, and the potential exposure of sensitive information to third-party tools via agents.


\subsubsection{Privacy Issues Caused by Automated Task Execution via LLM Agents}

Human instructions via prompts often contain ambiguities or omit crucial details. It is imperative to ensure the resilience and reliability of the agents' decisions, guaranteeing alignment between the actions performed by the agent and their intended tasks. Given that unknown risks lie in complex environments and user instructions, LLM agents are prone to causing unexpected privacy and safety issues \cite{xi2023rise, ruan2023identifying}. For instance, a LLM agent tasked to process emails might click on phishing links, leading to potential privacy breaches and financial loss. 
Similarly, if instructed to send an email with file content, the agent might accidentally include sensitive details like credit card information. 
In another example, \cite{fang2024llm} demonstrated that LLM agents could even 
exploit one-day vulnerabilities to hack websites based on task descriptions. 

However, due to their long-tail nature, identifying the risks associated with LLM agents is challenging. These risks arise as agents interact with various tools to execute tasks, coupled with the open-ended nature of potential issues and the substantial engineering effort required for testing such interactions.

\paragraph{\textbf{Mitigation}}

Several approaches assess whether user instructions are safe to execute without compromising user privacy and safety when evaluating LLMs' risk awareness regarding agent privacy.
ToolEmu \cite{ruan2023identifying} implemented a GPT-4 powered emulator with diverse tools and scenarios to identify potential failure modes in LLM agents and create sandbox states to trigger such failures. It also includes a GPT-4 powered safety evaluator to quantify risks. Similarly, AgentMonitor \cite{naihin2023testing} proposed using an LLM to monitor and halt unsafe actions, preventing potential safety and privacy issues on the open internet.
Both ToolEmu and AgentMonitor use LLMs to identify risky actions in agents. However, these methods have limitations: 
They often ignore core constraints, particularly in complex multi-turn interactions (back-and-forth interactions between LLM agents, users, and diverse environments), and rely on humans to provide risk descriptions.

To evaluate LLMs' risk awareness in agent safety through complex multi-turn interactions, Yuan et al. \cite{yuan2024r} developed R-Judge, a benchmark dataset. The dataset includes user instructions, agent action histories, and environment feedback annotated with safety labels. In evaluating automated harmful action detection, most LLMs struggled to identify safety and privacy risks, with GPT-4 achieving an F1 score of 72.52\% versus the human score of 89.07\%.

Hua et al. \cite{hua2024trustagent} presented TrustAgent, an agent framework employing three strategies to ensure safety: pre-planning (injecting safety knowledge before plan generation), in-planning (bolstering safety during plan generation), and post-planning (ensuring safety through post-planning inspection). Experiments show that TrustAgent enhances both safety and helpfulness. However, the study underscores the need for inherent reasoning abilities within LLMs to support truly safe agents.

Aside from the methods discussed that rely on external supervision from humans or other LLMs, which require significant investment in human labor and computational resources, alternative methods can operate without human supervision. Self-alignment \cite{sun2024principle} is an emerging paradigm where LLMs can independently achieve value alignment. Pang et al. \cite{pang2024self} explored this direction for achieving self-alignment of LLMs through multi-agent role-playing. This method allows the LLM to create a simulation environment that mirrors real-world multi-party interactions and simulates the social consequences of a user’s instruction. It records the textual interactions of the roles and summarizes them as the final output. Although the simulation process can be time-consuming during inference, fine-tuning can mitigate this issue.

\textit{
In summary, while these methods aim to prevent LLM agents from disclosing sensitive or personal data, they come with notable trade-offs and limitations tied directly to data privacy. Supervision-based approaches (ToolEmu, AgentMonitor, and TrustAgent) are effective at flagging and mitigating privacy risks but depend on human oversight. This reliance introduces its own privacy challenges (e.g., potential exposure of private information to supervisors) and can be resource-intensive. R-Judge seeks to automate the detection of privacy-related risks, yet it struggles with nuanced or context-dependent leaks, highlighting that even advanced models like GPT-4 still fall short of human-level detection in complex data-privacy scenarios. Meanwhile, self-alignment techniques attempt to reduce reliance on external supervision by enabling the model itself to learn protective behaviors around sensitive data. However, these methods are computationally expensive and have yet to prove robust across diverse real-world privacy contexts (where inadvertent disclosure can occur in unpredictable ways). Moving forward, designing scalable, adaptive frameworks with minimal human involvement remains critical: such approaches must specifically address how to protect sensitive information while ensuring that any oversight mechanisms themselves do not become new vectors for privacy compromise.}

\subsubsection{Adversarial Interactions of Agents}

Privacy risks can be further exacerbated by adversarial agents that mimic honest behavior and request additional private information under the guise of needing it for prompt execution tasks. Since LLM agents have demonstrated remarkable capabilities in generating human-like text, users may unwittingly disclose more information than they would to other service-providing systems \cite{zhang2023s}. In \cite{staab2023beyond}, the authors explore the interaction between LLM agents and users, where agents gather additional information through human-like conversations. 

LLM agents present complex privacy concerns due to their generative power and vulnerability to adversarial manipulation. Techniques such as ICL and RAG can be exploited to train agents for malicious purposes with minimal input. \cite{gu2024agent} found that modifying a single agent can rapidly lead to widespread harmful behaviors among agents without further intervention from the adversary. The nature of LLM agents and advanced techniques increases the risk of adversarial attacks and interactions. For evidence, \cite{tian2023evil} introduced Evil Geniuses, a virtual team that develops malevolent strategies and conducts Red-Blue exercises, demonstrating successful harmful actions through adversarial interactions. Similarly, \cite{struppek2024exploring} explored the potential for LLMs to act as adversaries by perturbing text samples to bypass safety measures. They investigated whether LLMs can inherently create adversarial examples from benign samples to deceive existing safeguards. Their findings underscore significant challenges for (semi-)autonomous systems using LLM agents, particularly concerning potential adversarial behavior.

\paragraph{\textbf{Mitigation}}

Chern et al. \cite{chern2024combating} proposed a multi-agent debate mechanism to improve quality and mitigate adversarial behavior. In this approach, agents self-evaluate through discussion and feedback. They tested this method with state-of-the-art models, evaluating susceptibility to red team attacks in single- and multi-agent settings. The results show that multi-agent debate generally produces less toxic responses to adversarial prompts without introducing new risks at inference time. However, the approach is resource-intensive, requires multiple queries, relies on knowledge from a single LLM, and does not address removing toxic outputs effectively.

Though multi-agent settings can mitigate this issue, multiple agents can have secret collusion between themselves. Motwani et al. \cite{motwani2024secret} formalized the issue of secret collusion in generative AI systems, proposing mitigation measures based on AI and security principles. They investigated monitoring the information content in agents' communications, focusing on agents' capabilities and incentives to evade detection. Using monitoring agents, they detected steganographic techniques that agents use to hide information by analyzing cover text anomalies and agent simulations. Future work could extend this by examining complexity and information theory, optimizing pressures, and exploring complex multi-agent settings.

Common alignment methods to reduce privacy risks 
include fine-tuning \cite{zeng2024autodefense, gururangan2020don} and reward modeling \cite{lee2023rlaif}. During fine-tuning, models are trained on specific datasets, incorporating human-generated examples to align with desired behaviors. Reward modeling involves optimizing a reward function to reflect desired outcomes, often using reinforcement learning techniques to adjust the model's behavior accordingly. Although intelligent personal agents should minimize user interruptions, integrating user opinions or human assistance can prove valuable when making significant decisions.

\textit{In summary, while multi-agent debate and alignment strategies enhance LLM privacy, they come with trade-offs. Debate-based methods reduce adversarial risks but are resource-intensive and knowledge-limited. Secret collusion raises concerns about hidden adversarial behavior, requiring costly monitoring. Fine-tuning and reward modeling enhance alignment but may not generalize well. 
}

\begin{figure*}[htbp]
    \centering
    \includegraphics[width=1\textwidth]{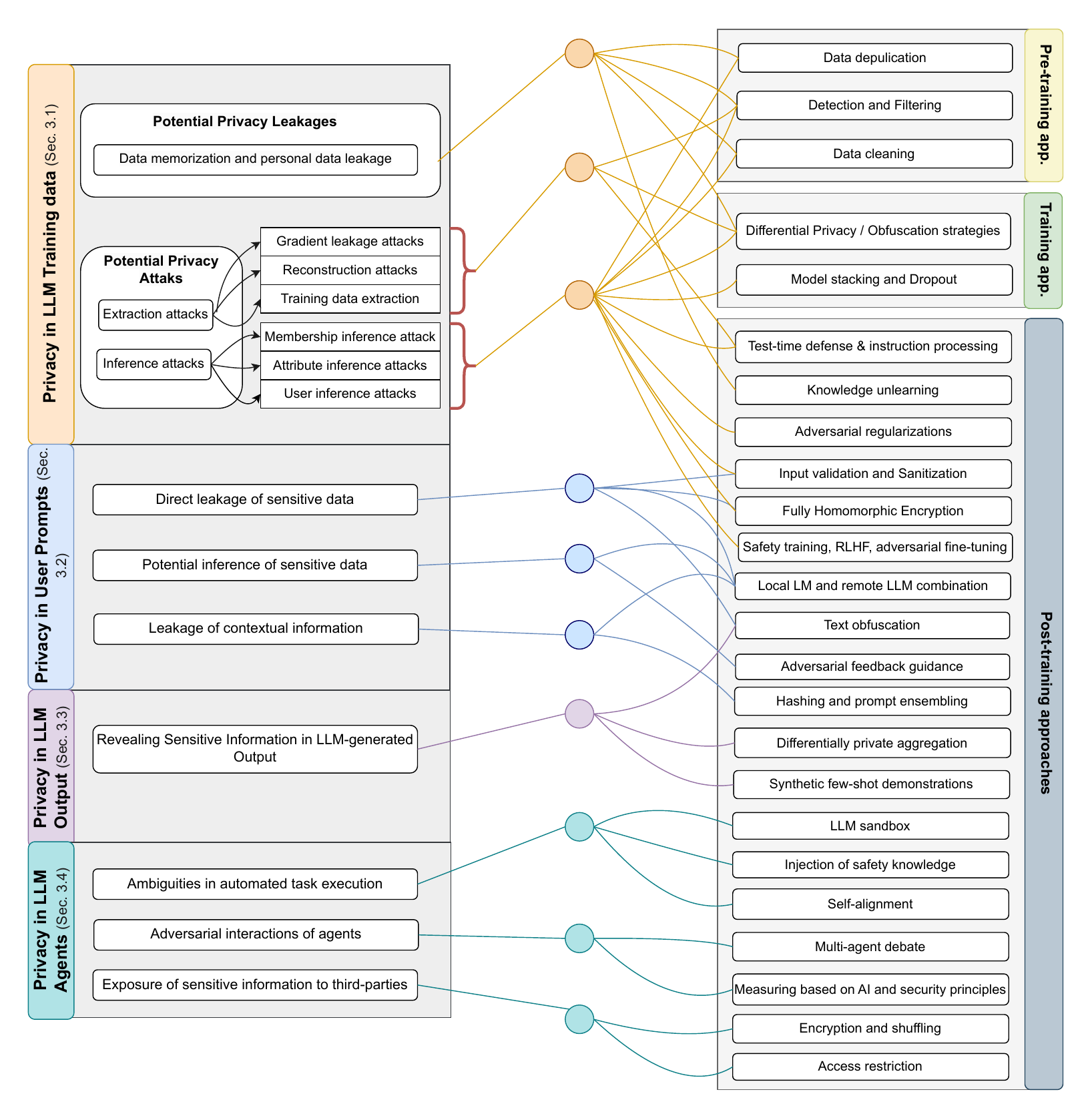}
    \caption{Summary of Privacy Challenges in LLM Systems and Mitigation Techniques in the Literature}
    \label{fig:summary}
    \Description{This figure provides a summary of the key privacy challenges faced by LLM systems along with the corresponding mitigation techniques proposed in the literature, highlighting how various approaches address different aspects of privacy risks.}
\end{figure*}

\subsubsection{Potential Exposure of Sensitive Information to Third-party Tools}

The next challenge is preventing the unnecessary and unauthorized disclosure of user-sensitive information during interactions with agents to third-party tools and the external world \cite{li2024personal}. 
The LLM agents orchestrate the task and invoke relevant third-party tools to execute it with the collected private information. However, in this process, the LLM possesses knowledge of the user’s private input and autonomously initiates queries in places without user awareness or inadvertently shares this information with the other agents or external tools.
Another threat model is adversarial third-party applications that can manipulate the context of interaction to trick LLM-based agents into exposing private information irrelevant to the task \cite{bagdasaryan2024air}. 

\paragraph{\textbf{Mitigation}}
 
The study in \cite{zhang2024privacyasst} examined privacy concerns for tool-using LLM agents. They proposed PrivacyAsst using encryption-based and shuffling-based solutions to preserve privacy. The encryption-based solution allows tasks to operate on encrypted inputs. In contrast, the shuffling-based solution uses attribute-based forgery generative models and an attribute shuffling mechanism to create privacy-preserving requests. However, limitations include the detectability of dummy prompts by advanced LLMs, the complexity and computational cost of encryption, reliance on third-party tools, and the burden of creating multiple dummy inputs on users.

AirGapAgent \cite{bagdasaryan2024air} focuses on preventing unintended data leakage during third-party interactions. It restricts access to only the necessary data for specific tasks, considering user privacy preferences. 
The design involves two LLMs: one minimizes data for sharing appropriately, while the other handles third-party interactions with the minimized data.
This ensures the agent can distinguish between private and non-private data for each task.

\textit{PrivacyAsst and AirGapAgent enhance privacy but face challenges: PrivacyAsst incurs high computational costs and third-party reliance, while AirGapAgent’s effectiveness depends on accurate filtering. Both highlight the trade-off between privacy and usability, requiring more scalable, adaptive solutions.}

\begin{tcolorbox}[colback=gray!15!white,colframe=gray!75!black,title=Key Takeaways]
LLM agents’ automated interactions with the physical and digital worlds
amplified their privacy vulnerabilities. 
Existing mitigation approaches often rely on human-guided LLM emulators to enforce user-specific privacy preferences and prevent sensitive information leakage.
Emerging approaches such as self-alignment seek to reduce human oversight by enabling agents to self-assess privacy risks through multi-agent role-playing.
\end{tcolorbox}

Overall, Figure \ref{fig:summary} provides an intricate portrayal of privacy concerns, encapsulating the four broader privacy challenges and the tailored mitigation technologies deployed at various phases of LLM systems, including data pre-training, training, and post-training approaches. 
A more detailed summary of these approaches can be found in Appendix: Table \ref{tab:summary}.

\section{Discussion and Future Directions}
\label{sec:disc}

This section highlights how some privacy-preserving methods could address multiple privacy threats and discusses their research challenges and potential future directions.

NER is widely used for direct leakage prevention in both training data and user prompts. 
Similarly, encryption and DP-based mechanisms can mitigate multiple privacy risks with trade-offs in processing and model performance.
Adversarial regularization, 
which enhances the robustness of LLM,
also plays a role in mitigating privacy attacks. 
However, these solutions are often complex due to fine-tuning between privacy preservation and model efficiency. 
Furthermore, many solutions operate in isolation, 
addressing individual threats rather than working together as part of a unified and adaptive privacy framework.

Privacy concerns in LLM systems usually stem from foundational models and extend to various downstream applications.
Traditional methods for protecting LLM training data fall short against more advanced attacks, especially when dealing with fine-tuned LLMs and their various ways of user interaction. 
These models are particularly susceptible to inference attacks
due to their complexity and natural language interaction. 
Existing mitigation techniques—such as detection, filtering, and model stacking— often come with trade-offs such as reduced performance, incomplete filtering, and higher latency and computational costs. 
To enhance LLM privacy, future approaches should explore hybrid models that integrate adaptive filtering, modular architectures, and human-in-the-loop validation. Reducing training data influence through prompt conditioning or modular architectures could provide additional privacy gains without sacrificing accuracy, particularly in fine-tuned models. Developing domain-specific privacy frameworks tailored for LLMs will also be crucial in aligning with emerging regulations. Future research should also focus on developing adaptive privacy mechanisms that dynamically assess conversation context and adjust LLM responses accordingly, ensuring that privacy-sensitive topics trigger enhanced protections such as real-time content filtering, obfuscation, or user alerts.

Other privacy-preserving approaches for generic models, such as cryptographic techniques, knowledge unlearning, and Federated Learning, 
show opportunities and challenges for the LLM domain. Cryptographic methods, while theoretically effective, are computationally expensive when applied to LLMs due to their large-scale processing requirements. 
Knowledge unlearning struggles with completely removing sensitive information when dealing with deeply integrated training data \cite{smith2023identifying}. 
Adapting such techniques for LLMs requires careful optimization to balance privacy, efficiency, and performance.


Another critical direction is the advancement of policy-driven AI governance. Efforts such as the EU AI Act~\cite{euai2024}, AI Safety Institutes~\cite{aisafetyinstituteInspect}, and Anthropic’s Constitutional AI \cite{bai2022constitutionalaiharmlessnessai} highlight the need for clearer guidelines on data collection, storage, and use in LLM systems. Automated auditing mechanisms and adversarial testing can further enhance transparency and accountability in LLM deployment, while continuous monitoring, adversarial testing, and proactive threat modeling could mitigate other privacy risks.


\section{Conclusion}
\label{sec:conc}

Our SoK identifies four main categories of privacy challenges: 
(i) privacy issues in LLM training data, 
(ii) privacy challenges associated with user prompts, 
(iii) privacy risks in LLM-generated outputs, 
and (iv) privacy vulnerabilities involving LLM agents.  
While the literature has extensively examined privacy issues in training data, 
privacy issues arising from user interactions and LLM outputs, particularly involving small and private datasets through fine-tuning, ICL, RAG, and agents, remain underexplored. 
Addressing these overlooked areas requires further research to strengthen privacy safeguards across user prompts, LLM outputs, and agent-based applications.

\bibliographystyle{ACM-Reference-Format}
\bibliography{main}
\section{Appendices}

\subsection{Source Selection and Strategy}
\label{app:source}

We conducted a literature review adhering to the Preferred Reporting Items for Systematic Reviews and Meta-Analyses (PRISMA) guidelines \cite{page2021prisma}. Our search spanned databases such as ACM, IEEE Xplore, Springer, ScienceDirect, and Google Scholar, targeting research papers that explore various privacy challenges in LLM systems published after 2022. 
The initial screening employed broad keywords such as `privacy challenges in LLM OR ChatGPT,' `privacy in LLM,' `privacy attacks in LLM,' `responsible LLM,' and `security AND privacy challenges in LLM.' We then refined our search strategy by categorizing the identified challenges (such as 'privacy in LLM agents', 'privacy in LLM prompts', `privacy in LLM output OR decision', and `privacy in LLM training data') via screening the identified challenges via initial search. Additionally, we used the mitigation methodologies found in the identified papers to search for further relevant studies. Survey papers on training data privacy challenges in LLMs were included as well.

\begin{figure}[ht]
    \centering
    \includegraphics[width=0.9\linewidth]{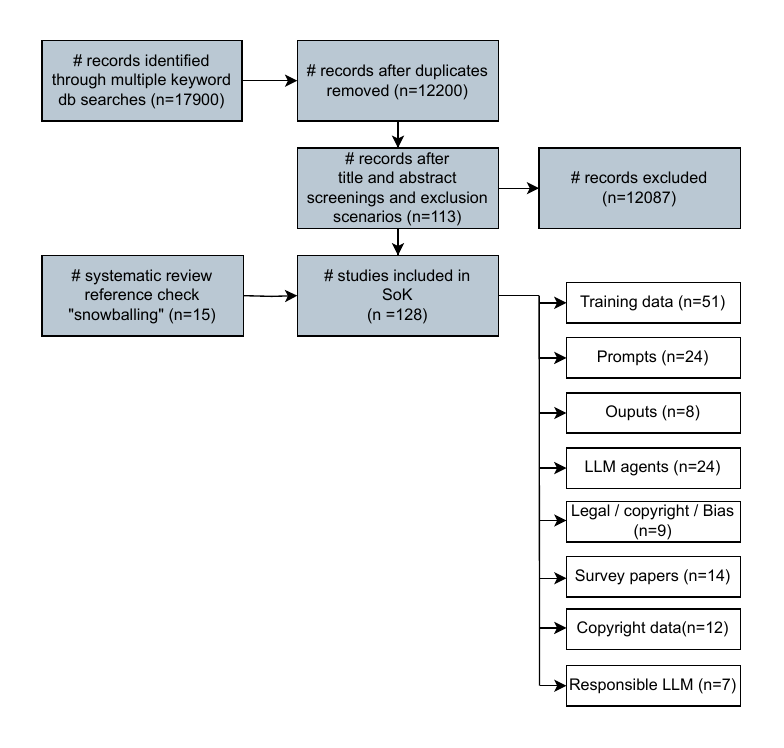}    
    \caption{PRISMA and Snowballing Approach for Paper Searching and Selection (\# implies number of)}
    \label{fig:source}
    \Description{This figure illustrates the paper selection process using the PRISMA methodology combined with a snowballing approach, outlining how relevant studies were systematically identified, screened, and included based on defined criteria.}
\end{figure}

Figure \ref{fig:source} shows the PRISMA and snowball approach of the paper search and selection process. Our search yielded a total of 17,900 records. We removed duplicates, screened titles and abstracts, and applied exclusion criteria, such as eliminating long-short repetitive papers, papers irrelevant to the focused area, security-related papers, those with no citations or published in non-A/A*/B conferences, and technical papers focused on root techniques with modifications. After this filtering, we were left with 101 records. We also employed a snowballing technique by reviewing the bibliographies of the identified papers to capture additional relevant studies. Ultimately, our study comprised 116 papers categorized as follows: Training data (n=51), Prompts (n=24), Outputs (n=8), LLM agents (n=24), Legal/copyright/bias (n=9), Survey papers (n=14), copyright data (n=12), and Responsible LLM (n=7).

\subsection{Glossary}
Table \ref{tab:glossary_table} lists the abbreviations used in the paper along with their descriptions.

\begin{table*}[!ht]
    \centering    
    \caption{The Glossary Table}
     \label{tab:glossary_table}   
    \begin{tabular}{p{4.8cm}|p{12.2cm}}
    \hline
    \rowcolor[HTML]{656565} 
        \color[HTML]{FFFFFF}\textbf{Abbreviations} & \color[HTML]{FFFFFF}\textbf{Description} \\ \hline
         Artificial Intelligence (AI) & The science of making machines that can think like humans\\ \hline
         Machine Learning (ML) & The capability of a machine to imitate intelligent human behavior\\ \hline
         Large Language Model (LLM) & Trained on vast datasets and characterized by extensive parameters to generate sensible responses \\ \hline         
         GDPR & General Data Protection Regulation \\  \hline
         HIPAA & Health Insurance Portability and Accountability Act of 1996 \\ \hline
         LLM systems user prompts & Natural language instructions to describe the task and achieve the desired outcomes\\ \hline
         LLM agents &  A system endowed with intricate reasoning capabilities, enabling it to interact with other systems and perform actions\\ \hline
         Personally Identifiable Information (PII) & Information that can be used to identify individuals \\ \hline
         Responsible AI (RAI) & An approach to developing and deploying AI in an ethical, trustworthy, safe, and legal way \\ \hline
         Fine-tuning in LLM & The process of adjusting the parameters of a pre-trained large language model to a specific task or domain \\ \hline
         In Context Learning (ICL) &  A method to adapt the LLM to a specific task by incorporating demonstrations into the prompt \\ \hline
         Retrieval Augmented Generation (RAG) & A technique where an LLM retrieves relevant information from external sources to enhance its generated responses\\ \hline
         Differential Privacy (DP) & A formal mathematical framework for ensuring privacy in data analysis. It provides strong guarantees that the inclusion or exclusion of any individual’s data in a dataset does not significantly impact the output of a computation, thereby protecting individual privacy. It achieves this by introducing controlled noise to statistical computations, making inferring private details about any single participant difficult. The strength of privacy protection is controlled by a parameter called epsilon ($\epsilon$)—a lower $\epsilon$ provides stronger privacy but may reduce accuracy. \\ \hline
         Knowledge unlearning & Knowledge unlearning is the process of selectively removing specific information from a trained model while preserving overall performance, which is crucial for privacy protection and regulatory compliance. Common approaches include exact unlearning (retraining without specific data), influence-based methods, and fine-tuning to mitigate privacy risks and data poisoning.\\ \hline
         Federated Learning (FL)  & A machine learning approach where models are trained across multiple decentralized devices or servers, preserving data privacy by keeping data local \\ \hline
         MIA & Membership Inference Attack: The objective of a MIA is to determine whether a specific data point $x$ (a sentence or document in LLM context) was included in the training dataset $\mathcal{D}$ of a model $M$. This is achieved by computing a membership score $f(x;M)$, which is then compared against a predefined threshold to decide whether the target sample was part of the training data by analyzing its output \cite{shokri2017membership}. 
         \\ \hline         
         Attribute Inference Attack & In an attribute inference attack, the adversary aims to predict sensitive attributes of an individual whose data was used to train a target model, given that the adversary already knows other attributes of that individual \cite{gong2018attribute}. \\ \hline
         Jailbreak attacks & A jailbreak attack is an adversarial technique used to bypass the safety restrictions and ethical guardrails of LLMs. These attacks manipulate the model into generating responses that it is designed to avoid, such as harmful, biased, or restricted content. 
         Jailbreak attacks exploit techniques like prompt engineering (deceptive inputs), role-playing (bypassing safeguards through scenarios), encoding tricks (obfuscation or special characters), and multi-turn exploits (gradual manipulation) to evade AI restrictions.\\ \hline
         RLHF & Reinforcement Learning from Human Feedback \\ \hline
         Intellectual Property (IP) & It encompasses creations of the mind, including inventions and works of art, protected by law through patents, copyrights, trademarks, and trade secrets \\ \hline
         API & Application Programming Interface \\ \hline
         NER & Named Entity Recognition: This involves identifying and classifying named entities in a text into predefined categories, such as the names of persons, organizations, locations, dates, times, and other entities. \\ \hline     
         
    \end{tabular}   
    
\end{table*}

\subsection{Responsible Large Language Models Concerns and Privacy Implications}
\label{app:resllm}

This section addresses responsible AI concerns in LLM systems, particularly from user interactions and LLM systems' inherent black-box nature. Responsible LLMs' concerns include ethical and legal issues due to LLMs tendency to copy proprietary material, transparency, reliability, explainability, accountability, fairness, and bias \cite{khowaja2023chatgpt, karamolegkou2023copyright}. 

The first challenge lies in the lack of transparency and interoperability of LLM, with service providers managing user data without clear mechanisms for users to understand its use. Ensuring stringent data protection is crucial due to the rising number of data breaches \footnote{https://www.statista.com/statistics/290525/cyber-crime-biggest-online-data-breaches-worldwide/} 
and the absence of strict regulations. The black-box nature of LLMs complicates verifying data processes, heightening privacy concerns. 


Another concern within LLM systems is the perpetuation of unfair discrimination and representational harm through reinforcing stereotypes and social biases \cite{khowaja2023chatgpt}. The association of social identities with LLM decisions often results in excluding specific social communities from the LLM's outcomes. Furthermore, hate speech generated by LLMs may incite violence and cause significant offense. Despite efforts by many LLM service providers to remove such content from their training corpora, bias and unfair discrimination remain persistent issues in LLMs.

Next, the potential for misinformation and disinformation is heightened by LLMs' ability to generate persuasive yet potentially false or misleading content. This capability can be exploited to spread misinformation, manipulate public opinion, or create fraudulent materials, posing significant risks to public health, democracy, and social harmony \cite{barman2024dark}.

Accountability in the deployment and use of LLM systems presents a critical issue, as it involves determining responsibility for the actions and outputs of these models \cite{jiao2024navigating}. When LLMs generate harmful or unethical content, such as misinformation or biased responses, it can be challenging to pinpoint who is responsible—the developers, users, or the organizations deploying the models. This issue is compounded by the complexity and opacity of LLMs, which can obscure the decision-making processes of the models. Establishing precise accountability mechanisms, such as transparent reporting standards, ethical guidelines for development, and oversight committees, is essential to ensure that all parties involved are held responsible for the impacts of LLM technologies \cite{barman2024beyond}.

\subsubsection{Mitigation}

Various approaches have been proposed to address responsible AI concerns, but they are still nascent for LLM systems. Potential solutions include improving user-friendly privacy policies and ethical data principles, conducting model audits, and using explainable AI techniques like Local Interpretable Model-agnostic Explanations (LIME) \cite{ribeiro2016should} and SHapley Additive exPlanations (SHAP) \cite{lundberg2017unified} to make LLM processes transparent. Feature visualization and activation maximization offer insights into a model's learning process, though their application to LLMs with millions of parameters is challenging. Additionally, it is crucial to educate users about the privacy implications of LLMs, their data usage policies, and the importance of protecting their privacy. Implementing mechanisms to quantify or explain privacy risks can further inform users of potential consequences. However, further work is needed to evaluate the effectiveness of these solutions.

In conclusion, addressing these concerns requires a multi-faceted approach that encompasses technical strategies (such as algorithms for detecting and mitigating biases), policy development (including privacy protection regulations), ethical guidelines (like respecting intellectual property), and stakeholder engagement (with collaboration among industry, academia, and regulatory bodies).

\newcommand{\tikzfully}{
    \begin{tikzpicture}[scale=0.5]
        \fill[fill=black] (0,0) circle (5pt);
    \end{tikzpicture}
}

\newcommand{\tikzhalf}{
\begin{tikzpicture}[scale=0.5]
    \fill[gray] (0,0) -- (90:5pt) arc[start angle=90, end angle=270, radius=5pt] -- cycle;
    \protect\draw (0,0) circle (5pt);
\end{tikzpicture}
}

\newcommand{\tikznot}{%
  \begin{tikzpicture}[scale=0.5]
    \protect\draw (0,0) circle (5pt);
  \end{tikzpicture}%
}

\subsection{Summary of Privacy Challenges and Solutions}

Table \ref{tab:summary} summarizes the privacy challenges and the details of existing mitigation techniques proposed in the literature.

\begin{table*}[!ht]     
    \centering
    \caption{Summary of Privacy Challenges in LLM Systems and Solutions. Experimented on `Open/closed source LLM system':\tikzfully = Closed-source LLM systems, \tikznot = Open source LLM systems, \tikzhalf = Both. `Threat model': \tikznot = Trusting LLM service provider and consider threats from external parties only, \tikzfully = Threat from both service provider and external parties. (This table continues on the next page.)}        
    \begin{tabular}{p{1.5cm}|p{2.1cm}|p{3.0cm}|p{6.2cm}|p{1.7cm}|p{1.4cm}}
    \hline
    \rowcolor[HTML]{656565} \color[HTML]{FFFFFF}
    \textbf{Privacy Challenges} & \color[HTML]{FFFFFF} \textbf{Specific Issues}& \color[HTML]{FFFFFF}\textbf{Solution} & \color[HTML]{FFFFFF}\textbf{Technical description} & \color[HTML]{FFFFFF}\textbf{Tested on Open/closed source LLM} & \color[HTML]{FFFFFF}\textbf{Threat model} \\ \hline

    \multirow{8}{1.5cm}{\begin{tikzpicture}\fill[trainingdata] (0,0) circle (5pt); \end{tikzpicture} In LLM Training data } &  \multirow{5}{1.5cm}{Data memorization and personal data leakage} & Data deduplication & Fuzzy deduplication techniques \cite{brown2020language} &  \begin{tikzpicture} \draw(0,0) circle (5pt); \end{tikzpicture} & \begin{tikzpicture} \draw (0,0) circle (5pt); \end{tikzpicture} \\ \cline{3-6}
    & & Detection and Filtering & Bloom filter \cite{ippolito2022preventing}, Filtering with restrictive terms \cite{ren2016recon}, PII scrubbing filters and NER \cite{sentryDataScrubbing}, Output filtering \cite{singh2024whispered} &  \begin{tikzpicture} \draw (0,0) circle (5pt); \end{tikzpicture} & \begin{tikzpicture} \draw (0,0) circle (5pt); \end{tikzpicture} \\ \cline{3-6}
    & & Data cleaning & Correcting errors and inconsistencies, implementing anonymization, data minimization, and secure practices to protect sensitive information \cite{yan2024protecting,brown2020language} &  \begin{tikzpicture} \draw(0,0) circle (5pt); \end{tikzpicture} & \begin{tikzpicture} \draw (0,0) circle (5pt); \end{tikzpicture} \\ \cline{3-6}
    & & Differential privacy & Adding noise to data when training\cite{zhao2022provably, behnia2022ew, zanella2020analyzing, singh2024whispered, li2021large, chang2023localization}  &  \begin{tikzpicture} \draw (0,0) circle (5pt); \end{tikzpicture} & \begin{tikzpicture} \draw (0,0) circle (5pt); \end{tikzpicture} \\ \cline{3-6}
    & & Knowlege unlearning & Force models to forget specific knowledge without requiring full retraining \cite{eldan2023whos, zhang2023composing, yao2023large} &  \begin{tikzpicture} \draw (0,0) circle (5pt); \end{tikzpicture} & \begin{tikzpicture} \draw (0,0) circle (5pt); \end{tikzpicture} \\\cline{2-6}

    & \multirow{5}{1.5cm}{Privacy attacks} & Model stacking / Dropout &  Model stacking: combines multiple base models\cite{du2024stacking}, Dropout: regularization technique used in neural networks to prevent overfitting \cite{xue2024repeat} &  \begin{tikzpicture} \draw (0,0) circle (5pt); \end{tikzpicture} & \begin{tikzpicture} \draw (0,0) circle (5pt); \end{tikzpicture} \\ \cline{3-6}
    &  & Test time defense and instruction processing &  Filters malicious inputs, detects abnormal queries, and post-processes LLM-generated output. \cite{robey2023smoothllm, li2022text} & \begin{tikzpicture}\fill[gray] (0,0) -- (90:5pt) arc[start angle=90, end angle=270, radius=5pt] -- cycle; \draw (0,0) circle (5pt); \end{tikzpicture}   & \begin{tikzpicture} \draw (0,0) circle (5pt); \end{tikzpicture} \\ \cline{3-6}
    &  & Adversarial regularizations & LLM training process employs robust optimization methods like adversarial training \cite{yoo2021towards} and robust fine-tuning \cite{dong2021should} to prevent malicious text attacks & \begin{tikzpicture} \draw (0,0) circle (5pt); \end{tikzpicture}  & \begin{tikzpicture} \draw (0,0) circle (5pt); \end{tikzpicture} \\ \cline{3-6}
    &  & Prompt-level approaches & Filtering adversarial prompts using rule-based detection or classifier-based approaches \cite{defending} & \begin{tikzpicture} \fill[gray] (0,0) -- (90:5pt) arc[start angle=90, end angle=270, radius=5pt] -- cycle;  \draw(0,0) circle (5pt); \end{tikzpicture} &  \begin{tikzpicture}  \fill (0,0) circle (5pt); \end{tikzpicture}  \\ \cline{3-6}
    & & Model-level approaches & Enhancing LLMs through safety training, RLHF, and adversarial fine-tuning \cite{jiang2025robustkv, bai2022constitutional} & \begin{tikzpicture}\draw (0,0) circle (5pt); \end{tikzpicture} &  \begin{tikzpicture}\draw (0,0) circle (5pt); \end{tikzpicture}  \\ \cline{3-6} \hline

    \multirow{8}{1.5cm}{\begin{tikzpicture}\fill[prompts] (0,0) circle (5pt);\end{tikzpicture} In Prompts }& \multirow{3}{1.8cm}{Direct leakage} & \multirow{3}{2.8cm}{Input validation and sanitization} & NER and encryption \cite{lin2024promptcrypt} & \begin{tikzpicture}\fill (0,0) circle (5pt);\end{tikzpicture} &  \begin{tikzpicture}  \fill (0,0) circle (5pt); \end{tikzpicture}   \\ \cline{4-6}
    & & & NER and Substitution/masking \cite{chen2023hide} & \begin{tikzpicture}\fill (0,0) circle (5pt);\end{tikzpicture} & \begin{tikzpicture}  \fill (0,0) circle (5pt); \end{tikzpicture} \\ \cline{4-6}
    & & & NER and obfuscation \cite{stracSecureSensitive} & \begin{tikzpicture}\fill[gray] (0,0) -- (90:5pt) arc[start angle=90, end angle=270, radius=5pt] -- cycle; \draw (0,0) circle (5pt); \end{tikzpicture} & \begin{tikzpicture}  \fill (0,0) circle (5pt); \end{tikzpicture} \\ \cline{3-6}
    & & Local small LM and remote LLM combination & Use NER to remove sensitive information before sending data to remote LLM, and add again locally\cite{hartmann2024can} & \begin{tikzpicture}\fill (0,0) circle (5pt);\end{tikzpicture} & \begin{tikzpicture}  \fill (0,0) circle (5pt); \end{tikzpicture} \\ \cline{3-6}
    & & Text obfuscation & Obscure original word information while retaining original word functionality \cite{zhou2023textobfuscator} & \begin{tikzpicture} \draw (0,0) circle (5pt); \end{tikzpicture} & \begin{tikzpicture}  \fill (0,0) circle (5pt); \end{tikzpicture} \\ \cline{3-6}
    & & Cryptographic approach: Fully harmonic encryption & Encrypting data in such a way that computations can be performed on the encrypted data without needing to decrypt it first \cite{ruoyan2025practical, zhang2024secpe, chen2022x, hao2022iron} & \begin{tikzpicture} \draw (0,0) circle (5pt); \end{tikzpicture} & \begin{tikzpicture}  \fill (0,0) circle (5pt); \end{tikzpicture} \\ \cline{2-6}
    & Inference of sensitive & Local small LM and remote LLM combination & Adversarial feedback-guided approach using LLM \cite{staab2024large} & \begin{tikzpicture}\fill[gray] (0,0) -- (90:5pt) arc[start angle=90, end angle=270, radius=5pt] -- cycle; \draw (0,0) circle (5pt); \end{tikzpicture}  & \begin{tikzpicture}  \fill (0,0) circle (5pt); \end{tikzpicture} \\ \cline{2-6}
    & \multirow{3}{1.9cm}{Contextual information leakage} & Hash operation & Transmitting only hashed values of sensitive data and service providers can revert back \cite{yim2023privacy} & - & \begin{tikzpicture} \draw (0,0) circle (5pt); \end{tikzpicture}  \\ \cline{3-6}
    & & Prompt ensembling & Using multiple different prompts to hide actual prompt of the user\cite{duan2023privacy} & \begin{tikzpicture}\draw (0,0) circle (5pt); \end{tikzpicture} & \begin{tikzpicture}  \fill (0,0) circle (5pt); \end{tikzpicture}  \\ \cline{3-6} 
    & & Local small LM and remote LLM combination & Remote LLMs provide high-level guidance and enhance the result in local LM with contextual information \cite{zhang2024cogenesis} & \begin{tikzpicture} \fill[gray] (0,0) -- (90:5pt) arc[start angle=90, end angle=270, radius=5pt] -- cycle;  \draw(0,0) circle (5pt); \end{tikzpicture} & \begin{tikzpicture}  \fill (0,0) circle (5pt); \end{tikzpicture} \\ \cline{3-6} \hline

    \end{tabular}
    \label{tab:summary}
\end{table*}

\begin{table*}[!ht]
    \centering    

    \begin{tabular}{p{1.5cm}|p{2.1cm}|p{3.0cm}|p{6.2cm}|p{1.7cm}|p{1.4cm}}
    \hline

    \multirow{4}{1.5cm}{\begin{tikzpicture}\fill[output] (0,0) circle (5pt);\end{tikzpicture} In LLM-generated Outputs } & \multirow{2}{1.8cm}{Revealing Sensitive Information} & Text obfuscation & Appending an another random text to the original prompt \cite{yao2024privacy} & - & \begin{tikzpicture}  \fill (0,0) circle (5pt); \end{tikzpicture} \\ \cline{3-6}
    & & Differentially private aggregation &  Aggregate and release responses without relying on any individual outputs \cite{wu2023privacy} & \begin{tikzpicture}\fill (0,0) circle (5pt);\end{tikzpicture} & \begin{tikzpicture}  \draw (0,0) circle (5pt); \end{tikzpicture} \\ \cline{4-6}
    & & &  Synthetic few-shot demonstrations \cite{tang2023privacy} & \begin{tikzpicture}\fill (0,0) circle (5pt);\end{tikzpicture} & \begin{tikzpicture}\fill (0,0) circle (5pt);\end{tikzpicture} \\ \cline{2-6}
    & Transparency in decision making & Explainable AI techniques & Feature importance analysis\cite{lundberg2017unified}, interactive exploration tools \cite{olah2018building}, and human-AI interpretation & - & - \\ \hline

    \multirow{7}{1.5cm}{\begin{tikzpicture}\fill[agents] (0,0) circle (5pt);\end{tikzpicture} In LLM Agents} & \multirow{3}{1.8cm}{Ambiguities in automated task execution} & LLM sandbox & LLM powered emulator and safety evaluator \cite{ruan2023identifying, naihin2023testing}, Complex multi-turn interactions \cite{yuan2024r} & \begin{tikzpicture}\fill[gray] (0,0) -- (90:5pt) arc[start angle=90, end angle=270, radius=5pt] -- cycle; \draw (0,0) circle (5pt); \end{tikzpicture} & \begin{tikzpicture}  \draw (0,0) circle (5pt); \end{tikzpicture} \\ \cline{3-6}
    & & Injecting safety knowledge in different stages & Safety knowledge injection in pre-planning, in-planning, and post-planning \cite{hua2024trustagent} & \begin{tikzpicture}\fill (0,0) circle (5pt);\end{tikzpicture} & \begin{tikzpicture}  \draw (0,0) circle (5pt); \end{tikzpicture} \\ \cline{3-6}
    & & Self-alignment & Self-alignment of LLMs through multi-agent role-playing \cite{pang2024self} & \begin{tikzpicture} \fill[gray] (0,0) -- (90:5pt) arc[start angle=90, end angle=270, radius=5pt] -- cycle;   \draw (0,0) circle (5pt);  \end{tikzpicture} & \begin{tikzpicture}  \draw (0,0) circle (5pt); \end{tikzpicture} \\ \cline{2-6}
    & \multirow{2}{1.8cm}{Adversarial interactions of agents} & Multi-agent debate mechanism & Self-evaluate through discussion and feedback & \begin{tikzpicture}  \draw (0,0) circle (5pt); \end{tikzpicture} & \begin{tikzpicture}  \draw (0,0) circle (5pt); \end{tikzpicture} \\ \cline{3-6}
    & & Mitigation measures based on AI and security principles & Monitoring agents' communication content \cite{motwani2024secret} & \begin{tikzpicture} \fill[gray] (0,0) -- (90:5pt) arc[start angle=90, end angle=270, radius=5pt] -- cycle;   \draw (0,0) circle (5pt);  \end{tikzpicture} & \begin{tikzpicture}  \fill (0,0) circle (5pt); \end{tikzpicture} \\ \cline{2-6}
    & \multirow{2}{1.8cm}{Exposure of Sensitive Information to External Tools} & Encryption and shuffling-based solutions & Operate on encrypted inputs and attribute-based forgery generative models with shuffling mechanism \cite{zhang2024privacyasst} & \begin{tikzpicture}  \fill (0,0) circle (5pt); \end{tikzpicture} & \begin{tikzpicture}  \fill (0,0) circle (5pt); \end{tikzpicture}\\ \cline{3-6}
    & & Access restriction methods & Limits access to necessary data based on user privacy preferences \cite{bagdasaryan2024air} & \begin{tikzpicture}  \fill[gray] (0,0) -- (90:5pt) arc[start angle=90, end angle=270, radius=5pt] -- cycle;  \draw (0,0) circle (5pt);   \end{tikzpicture} & \begin{tikzpicture}  \fill (0,0) circle (5pt); \end{tikzpicture} \\ \hline

    \end{tabular}
\end{table*}

\end{document}